%% file: main.tex
\documentclass[runningheads]{llncs}
\usepackage{booktabs}
\usepackage{url}
\usepackage{algorithm}
\usepackage{algorithmic}
\usepackage{amssymb,amsmath,array}
\usepackage{comment}
\usepackage{longtable}
\usepackage{color,graphicx}
\usepackage{float}
\usepackage{dblfloatfix}  
\usepackage{cite}

\begin{document}
\title{Explainable AI as a Social Microscope: A Case Study on Academic Performance}
\titlerunning{Explainable AI as a Social Microscope}
%

\author{Anahit~Sargsyan\inst{1,2} \and
        Areg~Karapetyan\inst{1,3} \and
        Wei~Lee~Woon\inst{1}
        \and~Aamena~Alshamsi\inst{1,4}} 
\institute{Department of Computer Science, Masdar Institute, Khalifa University, Abu Dhabi, UAE \and Division of Social Science, New York University Abu Dhabi, Abu Dhabi, UAE \and Research Institute for Mathematical Sciences (RIMS), Kyoto University, Kyoto 606-8502, Japan \and The MIT Media Lab, Massachusetts Institute of Technology, Cambridge, MA, USA}

\authorrunning{Sargsyan A., Karapetyan A., Woon W.L., Alshamsi A.}
%

%
\maketitle              
\begin{abstract}

Academic performance is perceived as a product of complex interactions between students' overall experience, personal characteristics and upbringing. Data science techniques, most commonly involving regression analysis and related approaches, serve as a viable means to explore this interplay. However, these tend to extract factors with wide-ranging impact, while \textit{overlooking variations specific to individual students}. Focusing on each student's peculiarities is generally impossible with thousands or even hundreds of subjects, yet data mining methods might prove effective in devising more targeted approaches. For instance, subjects with shared characteristics can be assigned to clusters, which can then be examined separately with machine learning algorithms, thereby providing a more nuanced view of the factors affecting individuals in a particular group. In this context, we introduce a data science workflow allowing for fine-grained analysis of academic performance correlates that captures the \textit{subtle differences in students' sensitivities to these factors}. Leveraging the Local Interpretable Model-Agnostic Explanations (LIME) algorithm from the toolbox of Explainable Artificial Intelligence (XAI) techniques, the proposed pipeline yields groups of students \textit{having similar academic attainment indicators}, rather than similar features (e.g. familial background) as typically practiced in prior studies. As a proof-of-concept case study, a rich longitudinal dataset is selected to evaluate the effectiveness of the proposed approach versus a standard regression model.


\keywords{Explainable AI  \and LIME \and Data Science \and Machine Learning \and Computational Social Science \and Academic Performance \and GPA Prediction.}
\end{abstract}

\input{intro.tex}

\input{ffcdataset.tex}
\input{methodology.tex}

\input{conclusion.tex}

\appendix
\input{append2}

\bibliographystyle{splncs04}
\bibliography{main}

\end{document}

%% file: intro.tex
\section{Introduction}

With far-ranging consequences on young people's lives and careers, academic performance is susceptible to various types and forms of influence. It is often path-dependent, correlating with an individual's past performance~\cite{coyle2008sat,salanova2010obstacles}. Furthermore, factors with significant impact on academic performance can be specific to the subject in question, such as intelligence and determination~\cite{colom2007fluid, tross2000not, pajares2001response}, or exogenous, resulting from the social, emotional and socioeconomic environment in which the individual was raised~\cite{pritchard2003using,graziano2007role,mcloyd1998socioeconomic}. Therefore, in general, investigation of academic performance predictors is attained through longitudinal studies~\cite{laidra2007personality, graziano2007role}.



Mainly, two directions are evidenced in this line of research. The first, followed in~\cite{jackson2006does, tillman2007family}, relies on statistical models to measure the correlation of a few variables that were premised on prior results in the literature. The second, attended in~~\cite{pandey2013decision, pal2013analysis, romero2010educational, asif2014predicting}, resorts to data science and machine learning techniques for extracting informative predictors from large datasets with thousands of candidate features.



While both approaches recognize the uniqueness of the students' backgrounds, the effect of the correlates is still determined as an aggregate over the entire study population/group, leaving the {\em subtle variations between individuals} largely overlooked. In particular, the former assumes that all the subjects are impacted by the same set of selected correlates in the same manner, whereas the latter seeks to derive a predictive model with high accuracy that generalizes to the overall population. However, no two {\em subjects are identical}, and factors profoundly affecting one individual might have a merely negligible impact on another person, even under comparable circumstances.

Against this backdrop, we introduce a novel data science approach in which (i) the predictors of academic performance for each student \textit{are identified and quantified} (ii) the study population is segmented into clusters based on the obtained values. The proposed pipeline allows the groups with \textit{similar success indicators} to be analyzed collectively, which should enable their effects to reinforce each other and be more readily discoverable. 


To quantify academic performance predictors specific to individual students, we avail of a recently developed XAI algorithm, known as LIME~\cite{ribeiro2016lime}. The outputs from LIME serve as ``explanations'', which are \textit{localized} in that a unique explanation is generated for each subject. Though descriptive on an individual case basis, these explanations are intrinsically disassociated, and thus their direct interpretation (one by one) becomes intractable with a growing number of subjects. This paper presents an efficient solution by grouping the students according to LIME coefficients, as detailed in Section~\ref{jorji}. Distinctively, under such clustering criterion, the subsequent analysis is explicitly centered at groups of students who \textit{share similar academic performance correlates}, as opposed to the classical approach of clustering in the feature space (i.e., based on observable characteristics such as gender, familial background or financial class). In a sense, with this scheme in place, it proves possible for subjects with fairly diverse backgrounds and needs to be grouped together, provided they share common markers of academic attainment.


As one demonstration, the proposed approach is benchmarked on a longitudinal dataset, released for the Fragile Families Challenge (FFC) competition, against a standard regression model. The results reveal a striking difference in the depth of insights gained, with the devised pipeline featuring prominently. While intended as a proof-of-concept, this preliminary study unveils findings on academic performance indicators that could serve social and data science communities. Furthermore, the workflow proposed herein can potentially pave the way towards more efficient and targeted intervention strategies by providing insights that would be inconceivable to achieve with traditional methods.

%% file: ffcdataset.tex
\section{Data and Pre-processing}\label{dataffc}


\subsection{FFC Dataset}
The examined dataset stems from the Fragile Families and Child Wellbeing study that documented the lives of over $4000$ births occurring between $1998$ and $2000$ in U.S. cities with at least $200,000$ population. As such, the study was carried out in the form of questionnaire surveys and interviews with parents shortly after the children's birth, and when the infants were $1$, $3$, $5$ and $9$ years old ({\it overall five waves}). The elicited data covered essential temporal information on the children's attitudes, parenting behavior, demographic characteristics, to name a few (further details of the dataset can be consulted in~\cite{REICHMAN2001303}). The data was released within the scope of FFC competition which sought to predict $6$ life outcomes, including Grade Point Average (GPA), based on these data records. Analysis of the overall results and ensuing findings of the competition are summarized in~\cite{Salganik8398}.

In total, the dataset comprises $4,242$ rows (one per child) and $12,943$ columns, including the unique numeric identifier. During FFC, however, only half of the data rows were released as a training set. Of these, $956$ entries had GPA values missing, and therefore the final dataset analyzed in this study totalled $1,165$ subjects, as appears in Fig.~\ref{fig:flowchart2} in the Appendix.

\subsection{Pre-processing and Feature Selection}
Before proceeding with this step, we remark that it is stipulated exclusively by nature (e.g., missing values) and properties (e.g., dimensionality) of the dataset under study and per se is not a principal constituent of the developed pipeline. Indeed, it is tailored specifically for the FFC dataset and might very well be substituted by any other appropriate routine yielding a sufficiently informative feature subset (i.e., \textit{with a decent predictive accuracy}) of reasonable cardinality. Thus, for clarity of exposition, the respective particulars are deferred to the Appendix.

\begin{figure}
\centering
  \includegraphics[scale=0.59]{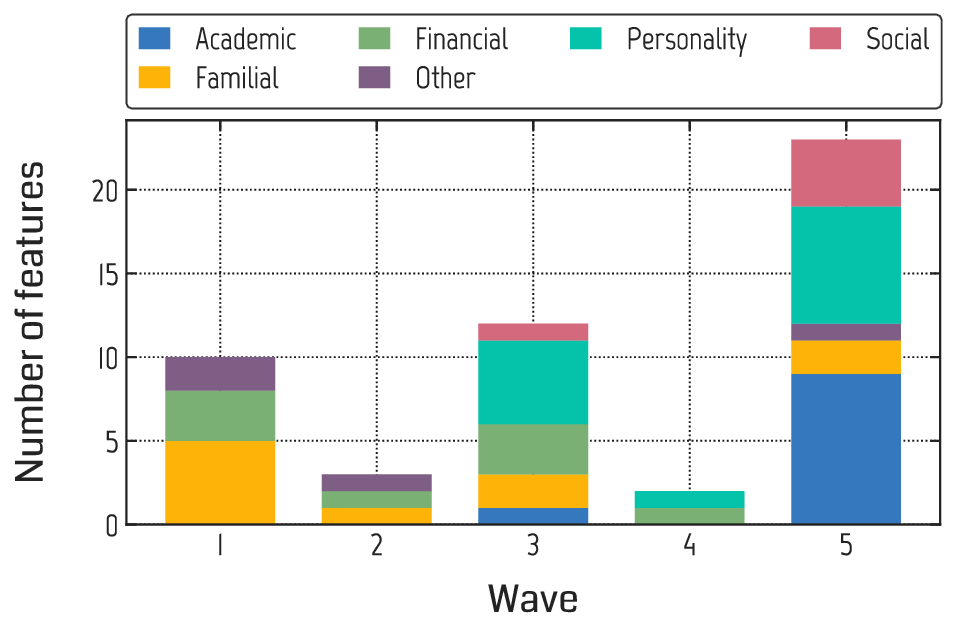}
  \caption{The distribution of selected $65$ features, categorized into $6$ major factor types, among the $5$ waves (i.e., over the course of the children's lives).}
  \label{fig:fancyness}
\end{figure}

The target subset of optimally descriptive features, as revealed through extensive experimentation and \textit{validated by its predictive accuracy}\footnote[1]{A mean squared error (MSE) of approximately $0.359$ was achieved under $3$-fold cross-validation (a result of comparable fidelity, submitted during the FFC, {\it secured a place in the top quartile} of the scoreboard).}, contained $65$ features, tabulated in Table~\ref{tab:selectedfeatures} in the Appendix. This pool of features forms the input for the proceeding analysis laid out in Section~\ref{jorji}. Figure~\ref{fig:fancyness} depicts the spread of these $65$ features across the $5$ waves (i.e., over the trajectory of children's lives). For each wave, the features are arranged into the following six categories: \{\textit{familial, financial, academic, social, personality, other}\} and their respective counts are illustrated in Fig~\ref{fig:fancyness} as a stacked bar chart.


As deduced from Fig.~\ref{fig:fancyness}, the distribution of features in familial and financial categories is skewed towards the early span of children's lives. For the correlates falling in academic, social and personality categories, the opposite trend is evidenced.   

%% file: methodology.tex
\vspace*{-30pt}
\section{Comparative Analysis}\label{jorji}

\begin{figure}
    \centering
    \includegraphics[scale=0.4]{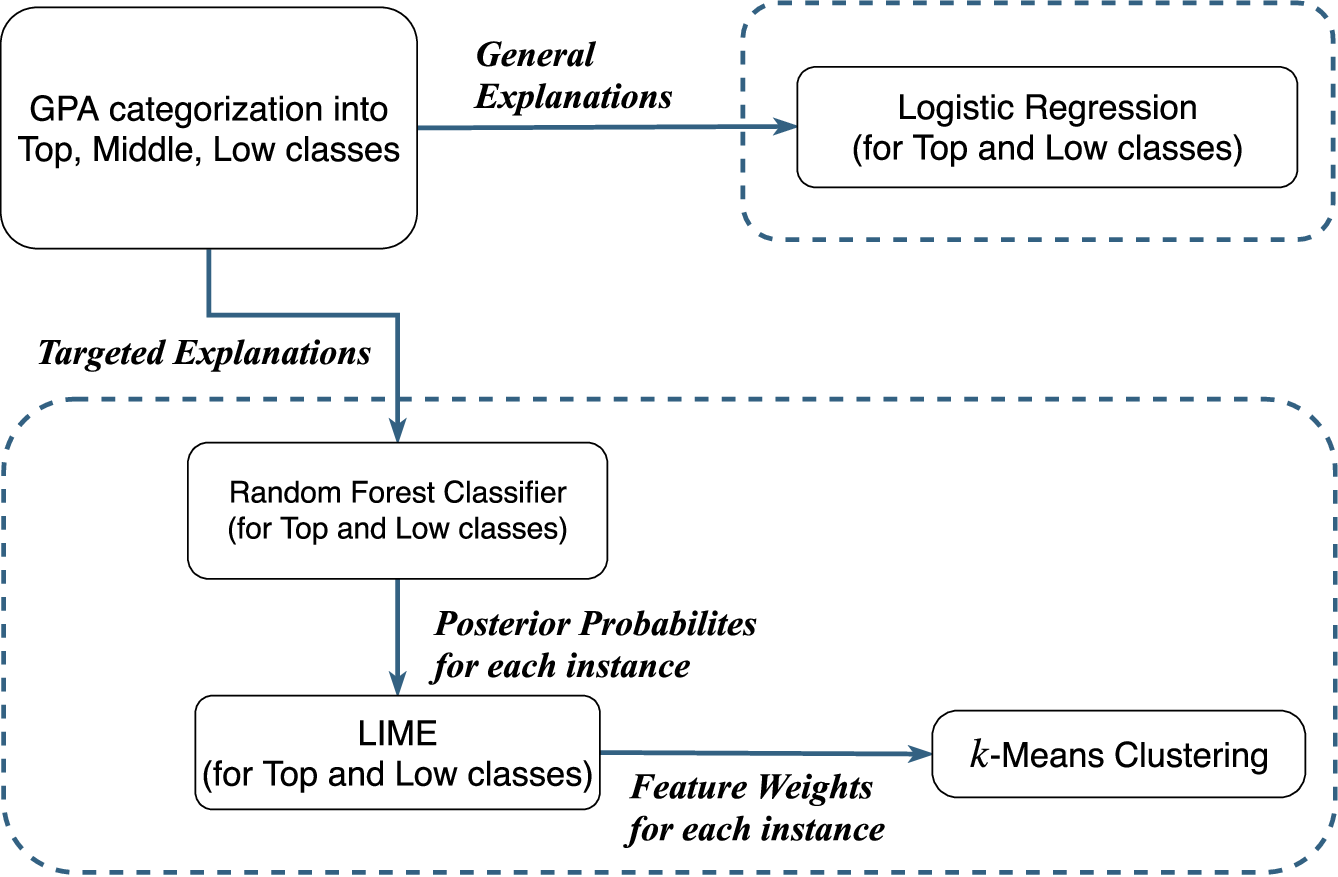}
\caption{Flowchart of the conducted comparative analysis including the featured methodology for obtaining targeted explanations.}
\label{fig:flowchart}
\end{figure}

This section contrasts the proposed approach against a conventional regression model, as illustrated in Fig.\ref{fig:flowchart}, and discusses the results.

We cast the problem as a classification task by discretizing the GPA scores into three classes, {\tt Low, Middle and Top}, defined respectively by the following ranges: $[1,2.5],(2.5,3.25),[3.25,4]$. Consequently, only the subjects falling into the {\tt Top} and {\tt Low} categories were retained ($861$ in total). The motivation is to steer the focus of classification algorithms towards the aspects discerning high and low performers. Indeed, the factors responsible for ``borderline'' performances are likely the ones with negligible impact, hence inferring them might obscure the results. On the other hand, omitting a large group of subjects could lead to the loss of pertinent data. Thus, the above thresholds were set according to the top and bottom 30\% percentiles of GPA score records. This ensured a solid number of participant students while retaining a sizable gap between the two classes.


\subsection{General Indicators}\label{vaxo}
\begin{figure}[ht!]
  \centering
    \includegraphics[scale=0.45]{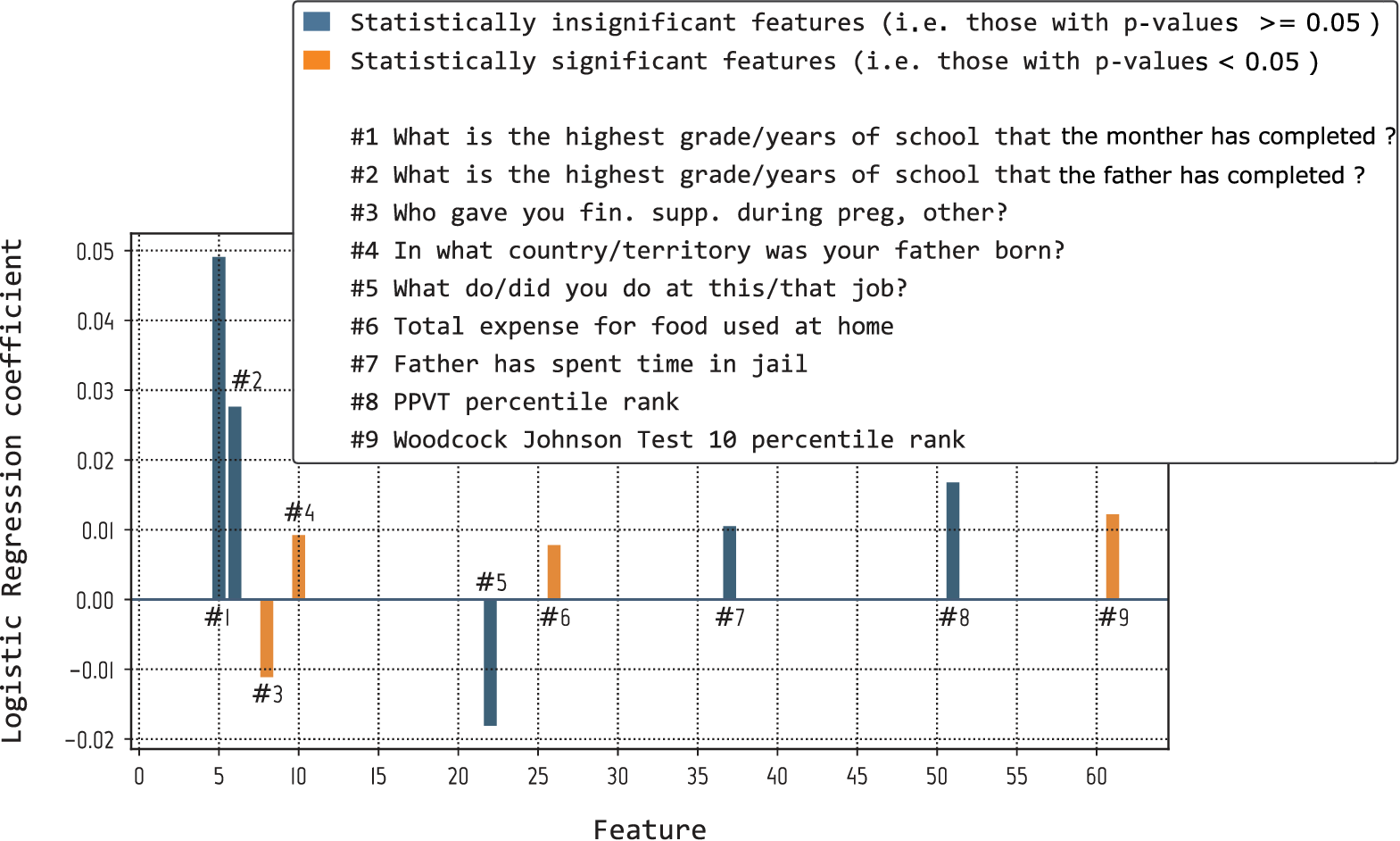}
  \caption{Logistic Regression coefficients of the selected $65$ features (ordered in a non-decreasing value of their waves) for {\tt Top} and {\tt Low} classes.} 
  \label{fig:LR_c_01}
\end{figure}

Following the common practice of previous works, in this initial phase, we employed the logistic regression algorithm to screen the selected features that broadly correlate with academic performance. The subjects from the {\tt Top} and {\tt Low} categories were fit to the model and the resulting coefficients, under L1 regularization, are presented in Fig.~\ref{fig:LR_c_01}.

As observed from Fig.~\ref{fig:LR_c_01}, test grades, along with other early metrics of academic performance, are imperative, and so are the factors associated with the child's social background. In particular, the indicators in familial and financial categories appear to influence children's academic performance predominantly at an early age. Whereas the correlates associated with scholastic aptitude manifest their effect mostly at later stages of subjects' lives. These observations are consistent with prior findings in the literature, providing some measure of validation. Overall, the most influential predictors are listed below.

\begin{enumerate}
\item The two most important factors relate to the parents' education~\cite{ermisch2001family}.
\item The \emph{Peabody Picture Vocabulary Test} (PPVT) percentile rank correlated with academic performance. PPVT is a standardized test designed to measure an individual's  vocabulary and comprehension and provide a quick estimate of verbal ability or scholastic aptitude. Another standard test's (known as Woodcock Johnson Test) percentile rank was identified as a significant indicator as well.
\item Interestingly, the fact that the father has been incarcerated\footnote[2]{Note that in Fig.~\ref{fig:LR_c_01} the positive correlation of this feature is due to the reversed order of values (i.e., the highest value indicates the father has not spent time in jail.)} - a proxy for family support - affects children's performance negatively. Contrariwise, the complexity/rank of the mother's job, a surrogate for the financial situation, conduced to enhanced academic performance.
\end{enumerate}

The emerging picture is compelling and multifaceted. On the one hand, test scores and academic aptitude occupy a central role, which is to be expected. Yet, there are indications that, beyond this, other features reflective of social and financial stability could also play a part, which strongly motivates the second, targeted part of this study. 

\subsection{Proposed Methodology: Targeted Indicators}



While the insights highlighted in Section~\ref{vaxo} were illuminating, they were extracted from the entire dataset, and the perspectives obtained were thus quite broad. To further extract targeted or localized indicators of academic success, we resorted to LIME~\cite{ribeiro2016lime}. For each instance, LIME produces a localized explanation of the classifier output by perturbing the feature values to generate a set of synthetic data points in the vicinity of the true instance. The posterior probability for each data point is estimated using the trained classifier, and a linear regression model is trained using the synthetic points as the inputs, and the posterior probabilities as the targets. The localized regression coefficients obtained in this way can then be interpreted as the {\em importance of a feature}, and are estimated separately for each subject.

\begin{figure}[ht!]
  \centering
  \includegraphics[scale=0.53]{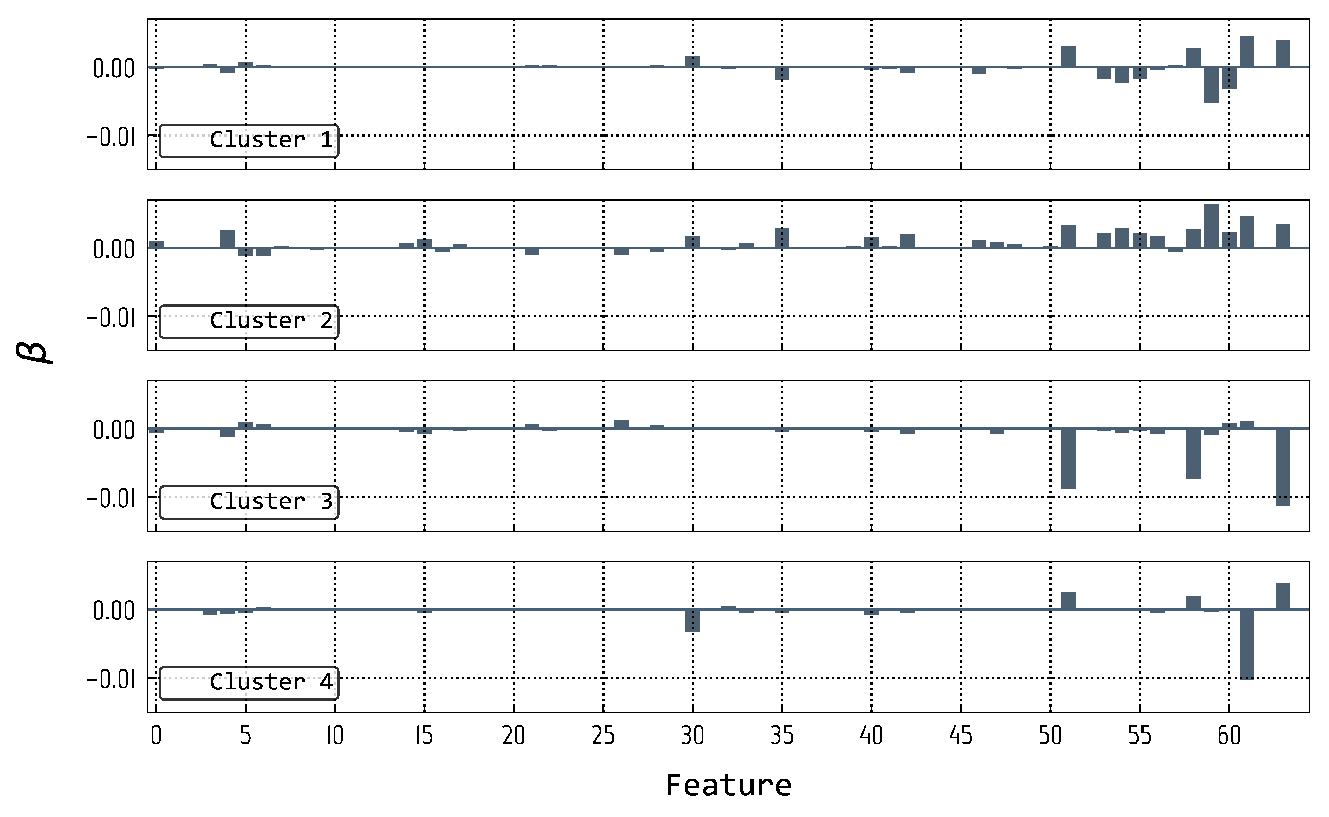}
  \caption{Clustering of students in {\tt Top} and {\tt Low} classes based on LIME coefficients. The horizontal axis represents the selected $65$ features sorted by their wave values in a non-decreasing order (i.e., over the course of children's lives). The vertical axis ($\beta$) depicts the deviations of the mean LIME coefficients for each cluster from the overall mean values of each LIME coefficient across the population. }
  \label{fig:LIME}
\end{figure}
This technique was adapted for the present context as follows. First, Random Forest classifier is trained on the data and is invoked to estimate the posterior probability for each instance. Then, LIME is applied to subjects falling into the {\tt Top} and {\tt Low} groups to produce feature weights specific to each subject, which are then clustered with the $k$-means algorithm. Each cluster is then characterized by the centroid of the LIME coefficients for the instances therein.

\subsection{Results and Discussion}
The $k$-means clustering results, with parameter $k=4$ (i.e., four clusters), appear in Fig.~\ref{fig:LIME}. This choice of $k$ enables a compact yet expressive representation of the underlying insights. In general, the higher the $k$, the more likely the groups are to resemble each other closely. On the other hand, when $k$ is small, one might possibly overlook factors critical to some subset of subjects.

In Fig.~\ref{fig:LIME}, to emphasize the differences between clusters, we focus on each feature's relative weights. That is, a cluster is represented in terms of the difference between its centroid and population means. As the figure suggests, the characteristics of the subjects vary significantly between the clusters. In particular, the salient patterns observed were as follows:


\begin{itemize}
\item \textbf{Cluster 1} (\textit{189 subjects}): The subjects in this cluster appear to be strongly influenced by features 51, 58, 61 and 63, which are all linked to test scores. Whereas features 59 and 60, related to the ability to pay attention, appeared to be less important.
\item \textbf{Cluster 2} (\textit{141 subjects}): This cluster was similar to Cluster 1 except that features 59 and 60 were more important than average.
\item \textbf{Cluster 3} (\textit{297 subjects}): In this cluster, features 51, 58 and 63 (all test score related), were all less important than average.
\item \textbf{Cluster 4} (\textit{234 subjects}): Here, feature 61 (the Woodcock Johnson Test score) was significantly less important, while features 51, 58 and 63 all had slightly stronger impacts on performance.
\end{itemize}
Overall, the features with values close to zero are the ones with a uniform effect on all individuals regardless of a cluster.

These results are deeply compelling in a number of ways. While Fig.~\ref{fig:LR_c_01} (results of the standard regression model) provided a broad overview of the overall success factors, the relative importance of each of these factors differed substantially among subjects. For example, in clusters 1 and 2, test scores appeared to have a substantial impact on future academic performance, while the opposite was observed in cluster 3. Also, features 59 and 60, which measure a student's general attentiveness, seemed relatively peripheral in cluster 1 and crucial in cluster 2 (and close to the average in clusters 3 and 4).



\begin{figure*}[!h]
  \centering
    \includegraphics[scale=0.53]{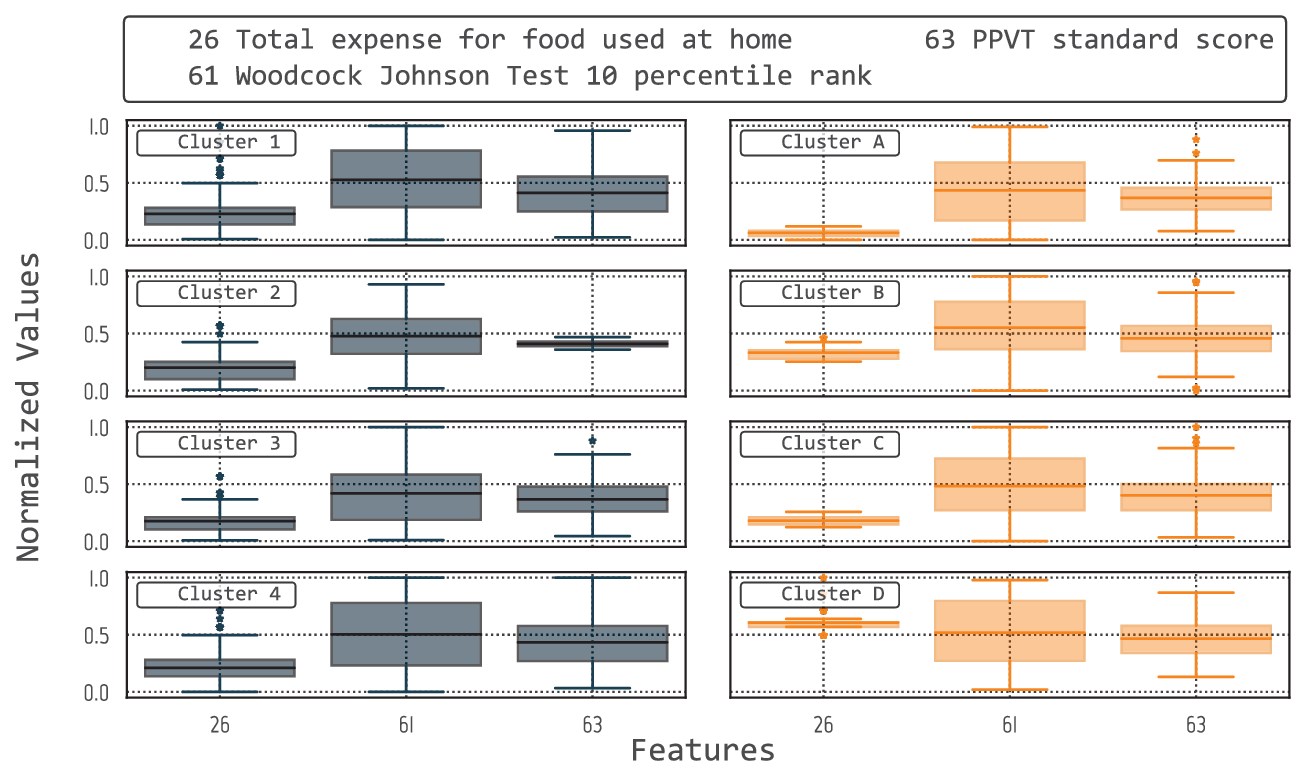}
\caption{Distributions of values of the selected numerical features within the clusters of students in {\tt Top} and {\tt Low} classes. Clusters $1$ to $4$ were produced with the proposed scheme (i.e., based on LIME coefficients), while clusters A to D with the conventional method (i.e., based on feature values).}
  \label{fig:new}
\end{figure*}

Another interesting observation was the spread of feature values \textit{within} clusters, which is depicted in Fig.~\ref{fig:new} (representing three selected numerical features). In particular, note the spread of values for feature 26, which is \textit{the total expense for food used at home}. Apparently, the feature values in the LIME clusters exhibit a far greater range compared to the conventional clusters, which tend to group people with similar financial situations. For the other features, the results are slightly less straightforward, but LIME clusters 3 and 4, in particular, also exhibit a much broader spread of values. This underscores that clustering with the LIME coefficients does not merely group students based on their personal or social circumstances, but rather in terms of the factors which affect their future academic performance.



%% file: conclusion.tex
\section{Future Work}

The present findings indicate the potential effectiveness of the proposed methodology in analyzing causal relationships in datasets alike FFC. However, this work was intended as a proof-of-concept case study, leaving open several avenues for future investigations. In particular, below listed are several promising directions.
\begin{enumerate}
    \item Analyze comparable data sets (e.g., The Millennium Cohort Study~\cite{LLCS410}), to test whether certain aspects of the observations are data-specific, or reflect true underlying patterns.
    \item Perform a thorough sensitivity analysis on the chosen subset of features as well as incorporate XAI techniques alternative to LIME (e.g., SHapley Additive exPlanations~\cite{NIPS2017_7062}).
    \item Apply the proposed approach to more diverse and larger datasets to demonstrate its generalizability to other use cases/study domains.
\end{enumerate}

\section{Concluding Remarks}

In this study, a novel data science pipeline is proposed, which conduces the identification of the specific features associated with academic performance in different groups of students. A clustering algorithm was employed to group these subjects, then targeted success indicators were extracted from each of these groups and scrutinized. We note that the present findings rely on a technique (LIME), which was developed relatively recently and should be treated as preliminary. However, if and when superior methods are proposed, they can similarly be incorporated into the devised workflow.

The key point is that such localized models are vital if we are to obtain a more nuanced view of the actual success indicators for specific children and families. The findings suggest that the children of fragile families can be given the best chance of success through interventions that are tailored to their individual needs, e.g. in some families, a small home loan could be the difference between a star student and a dropout, whereas in others a free mentoring scheme might turn more valuable.



%% file: append2.tex
\section*{Appendix}

\begin{figure}[ht!]
    \centering
    \includegraphics[scale=0.46]{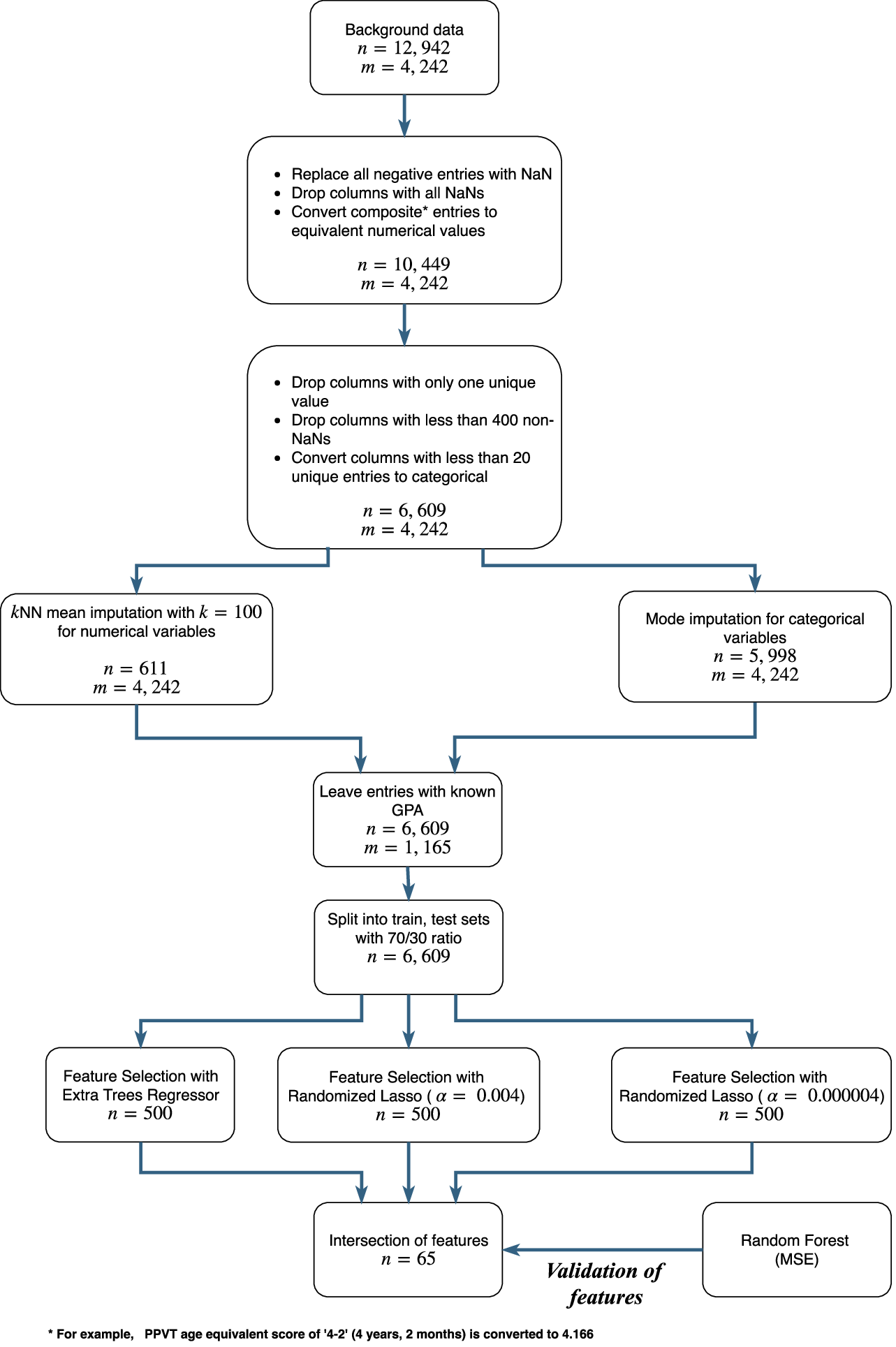}
\caption{A step-by-step illustration of the performed pre-processing and feature selection routines, depicted as a flowchart, with $n$ and $m$ standing for the number of features and entries in the given step, respectively.}
\label{fig:flowchart2}
\end{figure}

This section details the employed pre-processing and feature selection steps, portrayed as a flowchart in Fig.~\ref{fig:flowchart2}, along with the list of selected $65$ features.

\section{Details of the Pre-processing and Feature Selection Phase}

Due to the nature of the dataset under study, there were numerous instances of missing values where respondents either refused or were unavailable to answer. This was resolved through the judicious combination of data transformation, reduction, and imputation techniques, as listed below.

\begin{itemize}
    \item All missing and negative values were replaced by NaN and the columns with $0$ variance were removed. Then, only the columns having at least $400$ non-NaN values were retained.
   \item Features with less than $20$ unique values were treated as categorical and a simple median imputation was applied to replace the missing values.
    \item The variant of $k$NN ($k$-Nearest Neighbors) imputation algorithm, implemented in the Python package Fancyimpute, was leveraged, with value of $100$ for the parameter $k$, to estimate the remaining NaN values.
\end{itemize}

The number of features in the resulting dataset was reduced from over $12,900$ to $6,609$. However, this number of covariates was still exceedingly high, necessitating a subsequent feature selection phase. An array of filter- and wrapper-based methods, including Principal Component Analysis, Lasso, and Gradient Boosting Regression, were attempted in search of the most informative feature subset of reasonable cardinality. These methods were applied to the extracted pool of $6,609$ features, both recursively and explicitly, and probed under diverse parameter settings. The acquired subsets were then evaluated for their predictive accuracy across various models trained, effectively providing a means of validation. In essence, the latter step intends to establish the {\em overall validity of the model} underlying the proceeding analysis, thereby solidifying credibility of the explanations derived therein.

The target subset of optimally descriptive features was obtained by the following means. Feature importances were estimated by the Extra Trees Regressor algorithm (with $500$ estimators) and Randomized Lasso, and the top $500$ features were retained from each. For the latter, two different values were considered for the regularization parameter $\alpha$, namely $0.004$ and $0.000004$, thus resulting in two separate feature subsets. The intersection of these three subsets, containing $65$ features, led to maximized GPA prediction accuracy. In particular, with the Random Forest algorithm, an MSE of approximately $0.359$ was achieved under $3$-fold cross-validation.

\section{Selected Features}

\begin{longtable}{| p{0.1\textwidth}| p{0.65\textwidth}| p{0.18\textwidth}| p{0.01\textwidth}|}
\caption{The set of selected $65$ features listed in a non-decreasing order of their wave values.} \label{tab:selectedfeatures}\\
\cline{2-4}
 \multicolumn{1}{c|}{ }& \multicolumn{1}{c|}{\textbf{Feature name}} & \multicolumn{1}{c|}{\textbf{Respondent}} & \multicolumn{1}{c|}{\textbf{Wave}} \\ \hline
\endfirsthead
\endhead
\multicolumn{1}{|l|}{0} & Is the home/apartment were you currently reside owned/rented? & mother &1 \\
\multicolumn{1}{|l|}{1} & Who gave you fin. supp. during preg, (BF) family?	& mother &1\\
\multicolumn{1}{|l|}{2} & Were you living with both of your bio parents at age 15? & mother &	1\\
\multicolumn{1}{|l|}{3} & Father baseline education (combined report)& father &	1\\
\multicolumn{1}{|l|}{4} & Are BM \& BF living together?& father &	1\\
\multicolumn{1}{|l|}{5} & What is the highest grade/years of school that you have completed?& mother&	1\\
\multicolumn{1}{|l|}{6} & What is the highest grade/years of school that BF have completed?& mother &	1\\
\multicolumn{1}{|l|}{7} & Are you and BM living together now?& father &	1\\
\multicolumn{1}{|l|}{8} & Who gave you fin. supp. during preg, other?&mother &	1\\
\multicolumn{1}{|l|}{9} & How imp for successful marriage, wife has steady job?& mother&	1\\
\multicolumn{1}{|l|}{10} & In what country/territory was your father born?& father &	2\\
\multicolumn{1}{|l|}{11} & Could you count on someone to co-sign for a loan for \$1000? & mother& 2\\
\multicolumn{1}{|l|}{12} & Could you count on someone to co-sign for a loan for \$5000?& mother&	2\\
\multicolumn{1}{|l|}{13} & Have you ever been in program to help relationship?	& father &3\\
\multicolumn{1}{|l|}{14} & He/she hits others& home visit&	3\\
\multicolumn{1}{|l|}{15} & Is mother living with father or current partner? & mother&	3\\
\multicolumn{1}{|l|}{16} & There are quite a few things that bother you about your life?& home visit&	3\\
\multicolumn{1}{|l|}{17} & (He/she) can't sit still, is restless or hyperactive	& home visit & 3\\
\multicolumn{1}{|l|}{18} & Is there someone to co-sign for a bank loan with you for \$1,000?	& mother & 3\\
\multicolumn{1}{|l|}{19} & What about co-signing for \$5,000?	& mother &3\\
\multicolumn{1}{|l|}{20} & Does father have any other children by someone else?& mother &	3\\
\multicolumn{1}{|l|}{21} & Do mother and father currently live together?	& home visit &3\\
\multicolumn{1}{|l|}{22} & What do/did you do at this/that job?& father &	3\\
\multicolumn{1}{|l|}{23} & Did father see child more than once in the past 30 days?& father &	3\\
\multicolumn{1}{|l|}{24} & In past year, did you think you were eligible for welfare at any time?	& mother &3\\
\multicolumn{1}{|l|}{25} & (He/she) feels (he/she) has to be perfect	& home visit&4\\
\multicolumn{1}{|l|}{26} & Total expense for food used at home	&home visit &4\\
\multicolumn{1}{|l|}{27} & Who gave help: other relatives of father& father&	4\\
\multicolumn{1}{|l|}{28} & You often have the feeling that you cannot handle things very well& home visit&	4\\
\multicolumn{1}{|l|}{29} & Did (you/a family member living w you) use government food stamps, last month& home visit&	4\\
\multicolumn{1}{|l|}{30} & Woodcock Johnson Test 10 age equivalency& home visit&	5\\
\multicolumn{1}{|l|}{31} & You could ask child's mother's parents/relatives for help/advice&father &	5\\
\multicolumn{1}{|l|}{32} & Feature f5g351\footnote[3]{See https://codalab.fragilefamilieschallenge.org/f/api/codebook/ for the
description}& father&	5\\
\multicolumn{1}{|l|}{33} & I get in trouble for talking and disturbing others& kid &	5\\
\multicolumn{1}{|l|}{34} & Child shows off or clowns& primary caregiver &	5\\
\multicolumn{1}{|l|}{35} & Child's science and social studies	&teacher &5\\
\multicolumn{1}{|l|}{36} & Child has participated in Title I ESL/bilingual& teacher &	5\\
\multicolumn{1}{|l|}{37} & Father has spent any time in jail& mother&	5\\
\multicolumn{1}{|l|}{38} & Child repeated 4th grade& primary caregiver &	5\\
\multicolumn{1}{|l|}{39} & You could ask friends/neighbors/co-workers for help/advice& mother &	5\\
\multicolumn{1}{|l|}{40} & Child is disobedient at home	&primary caregiver &5\\
\multicolumn{1}{|l|}{41} & Child keeps desk clean and neat without being reminded& teacher &	5\\
\multicolumn{1}{|l|}{42} & Child is inattentive or easily distracted	& primary caregiver&5\\
\multicolumn{1}{|l|}{43} & It's hard for me to pay attention	& kid&5\\
\multicolumn{1}{|l|}{44} & You could ask child's mother for help/advice	& father&5\\
\multicolumn{1}{|l|}{45} & Child can't concentrate, can't pay attention for long& primary caregiver &	5\\
\multicolumn{1}{|l|}{46} & Child's language and literacy skills& teacher&	5\\
\multicolumn{1}{|l|}{47} & It's hard for me to finish my schoolwork & kid &	5\\
\multicolumn{1}{|l|}{48} & I worry about doing well in school& kid&	5\\
\multicolumn{1}{|l|}{49} & I worry about taking tests&kid &	5\\
\multicolumn{1}{|l|}{50} & Number of families on block know well& primary caregiver&	5\\
\multicolumn{1}{|l|}{51} & PPVT percentile rank&home visit &	5\\
\multicolumn{1}{|l|}{52} & Child shows anxiety about being with a group of children& teacher &	5\\
\multicolumn{1}{|l|}{53} & Child finishes class assignments with time limits& teacher&	5\\
\multicolumn{1}{|l|}{54} & Child uses free time in an acceptable way	& teacher&5\\
\multicolumn{1}{|l|}{55} & Child says nice things about self/others when appropriate	&teacher &5\\
\multicolumn{1}{|l|}{56} & Who usually initiated the contact& primary caregiver &	5\\
\multicolumn{1}{|l|}{57} & Number of child's close friends you& primary caregiver &	5\\
\multicolumn{1}{|l|}{58} & PPVT age equivalency& home visit&	5\\
\multicolumn{1}{|l|}{59} & Child attends to your instructions& teacher &	5\\
\multicolumn{1}{|l|}{60} & Child ignores peer distractions when doing class work& teacher &	5\\
\multicolumn{1}{|l|}{61} & Woodcock Johnson Test 10 percentile rank& home visit&	5\\
\multicolumn{1}{|l|}{62} & Frequency kids picked on you or said mean things to you & kid&	5\\
\multicolumn{1}{|l|}{63} & PPVT standard score& home visit &	5\\
\multicolumn{1}{|l|}{64} & Being a parent is harder than I thought it would be& father &	5\\
\hline
\end{longtable}